\documentclass{ws-ijmpd}
\usepackage{cite}
\usepackage{graphics}
\usepackage{graphicx}
\usepackage{tabularx}
\usepackage{makeidx}
\usepackage{epsfig}
\usepackage[caption=false]{subfig}
\usepackage{colortbl}
\usepackage{colordvi}
\usepackage{verbatim}
\usepackage{amsmath}
\usepackage{enumerate}
\usepackage{hyperref}
\usepackage{dcolumn}
\usepackage{bm}
\usepackage{amssymb,mathrsfs}
\usepackage{ulem}

\def\pa{{\partial}}
\def\l{\left}
\def\r{\right}
\def\d{{\rm d}}
\def\f{\frac}

\def\nn{\nonumber} 
\def\pa{{\partial}}
\def\f{\frac}
\def\l{\left}
\def\r{\right}
\def\d{{\rm d}}
\def\Mpl{M_{_{\rm Pl}}}
\def\beq{\begin{equation}}
\def\eeq{\end{equation}} 
\def\beqa{\begin{eqnarray}}
\def\eeqa{\end{eqnarray}}

\def\cN{\mathcal N}

\def\nn{\nonumber} 
\def\pa{{\partial}}
\def\f{\frac}
\def\l{\left}
\def\r{\right}
\def\d{{\rm d}}
\def\k{{\bm k}}
\def\x{{\bm x}}
\def\R{{\rm {R}}}
\def\I{{\rm {I}}}

\def\cN{\mathcal N}
\def\Mpl{M_{_{\rm Pl}}}

\newcommand{\viz}{\textit{viz.~}}
\newcommand{\ie}{\textit{i.e.~}}

\begin{document}

\markboth{D. Jaffino Stargen, V. Sreenath and L. Sriramkumar}
{Quantum-to-classical transition and imprints of CSL in classical bouncing universes} 
%
\catchline{}{}{}{}{}
%

%
%
%
%



\title{Quantum-to-classical transition and imprints of continuous\\
spontaneous localization in classical bouncing universes}
\author{D. Jaffino Stargen\footnote{Current address:~Indian Institute of 
Science Education and Research Mohali, Knowledge city, SAS Nagar, 
Punjab 140306, India. E-mail:~jaffino@iisermohali.ac.in}}
\address{Department of Physics, Indian Institute of Technology Madras, 
Chennai~600036, India}
\author{V.~Sreenath\footnote{Current address: Department of Physics, 
National Institute of Technology Karnataka, Surathkal,
Mangalore 575025, India. E-mail:~sreenath@nitk.edu.in}}
\address{Department of Physics and Astronomy, Louisiana State University, 
Baton Rouge, LA~70803, U.S.A.}
\author{L.~Sriramkumar\footnote{E-mail:~sriram@physics.iitm.ac.in}}
\address{Department of Physics, Indian Institute of Technology Madras, 
Chennai~600036, India}
\maketitle
\begin{abstract}
The perturbations in the early universe are generated as a 
result of the interplay between quantum field theory and gravitation. 
Since these primordial perturbations lead to the anisotropies in the 
cosmic microwave background and eventually to the inhomogeneities in 
the Large Scale Structure (LSS), they provide a unique opportunity to 
probe issues which are fundamental to our understanding of quantum 
physics and gravitation. 
One such fundamental issue that remains to be satisfactorily addressed 
is the transition of the primordial perturbations from their quantum 
origins to the LSS which can be characterized completely in terms of 
classical quantities.  
Classical bouncing universes provide an alternative to 
the more conventional inflationary paradigm as they can help overcome 
the horizon problem in a fashion very similar to inflation.
While the problem of the quantum-to-classical transition of the primordial 
perturbations has been investigated extensively in the context of 
inflation, we find that there has been a rather limited effort towards
studying the issue in classical bouncing universes.
In this work, we analyze certain aspects of this problem with the example
of tensor perturbations produced in classical matter and near-matter
bouncing universes.
We investigate the issue mainly from two perspectives.
Firstly, we approach the problem by examining the extent of squeezing 
of a quantum state associated with the tensor perturbations with the
help of the Wigner function. 
Secondly, we analyze the issue from the perspective of the quantum 
measurement problem. 
In particular, we study the effects of wave function collapse, using 
a phenomenological model known as continuous spontaneous localization, 
on the tensor power spectra.
We conclude with a discussion of results.
\end{abstract}


%
%


\section{Introduction}

The current cosmological observations seem to be well described by 
the so-called standard model of cosmology, which consists of the
$\Lambda$CDM model, supplemented by the inflationary 
paradigm~\cite{planck-2015-ccp,planck-2015-ci}. 
The primary role played by inflation is to provide a causal mechanism 
for the generation of the primordial perturbations~\cite{i-reviews}, 
which later lead to the anisotropies in the Cosmic Microwave Background 
(CMB) and eventually to the inhomogeneities in the Large Scale Structure 
(LSS)~\cite{cosmology-texts}. 
The nearly scale invariant power spectrum of primordial perturbations
predicted by inflation has been corroborated by the state of the art
observations of the CMB anisotropies by the Planck 
mission~\cite{planck-2015-ci}.
Despite the fact that inflation has been successful in helping to
overcome some of the problems faced within the hot big bang model, 
the issue of the big bang singularity still remains to be addressed. 
Moreover, the remarkable efficiency of the inflationary scenario has led
to a situation wherein, despite the constant improvement in the accuracy
and precision of the cosmological observations, there seem to exist 
too many inflationary models that remain consistent with the 
data~\cite{planck-2015-ci}.
This situation has even provoked the question of whether, as a 
paradigm, inflation can be falsified at all (in this context, see the 
popular articles~\cite{pa}).
Due to these reasons, it seems important, even imperative, to 
systematically explore alternatives to inflation.
One such alternative that has drawn a lot of attention in the literature
are the classical bouncing scenarios~\cite{bs-reviews}.

\par
In bouncing models, the universe goes through an initial phase of contraction, 
until the scale factor reaches a minimum value, and it undergoes expansion 
thereafter~\cite{bs-reviews}. 
Driving a bounce often requires one to violate the null energy condition and 
hence, unlike inflation, they cannot be driven by simple, canonical scalar fields.
In fact, the exact content of the universe which is responsible for the bounce 
remains to be satisfactorily understood. 
Also, concerns may arise whether quantum gravitational effects can become 
important at the bounce~\cite{bounces-in-lqg}. 
To avoid such concerns, one often considers completely classical bounces wherein 
the energy densities of the matter fields driving the bounce always remain much
smaller than the Planckian energy densities.
In a fashion similar to slow roll inflation, certain bouncing models referred
to as near-matter bounces, can also generate nearly scale invariant power 
spectra~\cite{nsis-in-b,mb,rathul-2018,rathul-2019}, as is demanded by the 
observations~\cite{planck-2015-ccp,planck-2015-ci}. 
However, while proposing an inflationary model seems to be a rather easy task
(which is reflected in the multitude of such models), a variety of problems
(such as the need for fine tuned initial conditions and the rapid growth of 
anisotropies, to name just two) plague the bouncing models~\cite{bs-reviews}.
It would be fair to say that a satisfactory classical bouncing scenario that 
is devoid of these various issues is yet to be constructed. 

\par

The generation of primordial perturbations in the early universe, whether 
in a bouncing or in an inflationary scenario, is a result of an interplay 
between quantum and gravitational physics~\cite{dcp,squeezing,jerome-2008}. 
Since it is the quantum perturbations that lead to anisotropies in the CMB 
and inhomogeneities in the LSS, it provides a unique window to probe 
fundamental issues pertaining to quantum and gravitational physics. 
One such issue of interest is the mechanism underlying the transition of 
the quantum perturbations generated in the early universe to the LSS that 
can be completely described in terms of correlations involving classical
stochastic variables, in other words, the quantum-to-classical transition
of the primordial perturbations.
 
\par

While the issue of the quantum-to-classical transition of primordial
perturbations has been studied to a good extent in inflation~\cite{dcp,
squeezing,jerome-2008}, we find that there has been hardly any effort 
in this direction in the context of bouncing scenarios (see, however, 
Ref.~\cite{leon-2016} which addresses issues similar to what we shall
consider here).
In this work, we shall investigate the problem in the case of tensor
perturbations produced in a class of classical bouncing scenarios.
We shall approach the problem from two different perspectives. 
Firstly, we shall examine the extent of squeezing of the quantum state
associated with the tensor perturbations using the Wigner function~\cite{dcp}. 
It has been found that, in the context of inflation, the primordial quantum 
perturbations become strongly squeezed once the modes leave the Hubble
radius~\cite{squeezing,jerome-2008}. 
In strongly squeezed states, the quantum expectation values can be
indistinguishable from classical stochastic averages of the correlation
functions, such as those used to characterize the anisotropies in the 
CMB and the LSS~\cite{dcp}.
Specifically, we shall investigate if the Wigner function and the parameter
describing the extent of squeezing behave in a similar manner in the 
bouncing scenarios.

\par

Secondly, we shall study the issue from the perspective of a quantum 
measurement problem.
The quantum measurement problem concerns the phenomenon by which a 
quantum state upon measurement collapses to one of the eigenstates 
of the observable under measurement. 
In the cosmological context, this problem translates as to how the quantum 
state of the primordial perturbations collapses into the eigenstate, say, 
corresponding to the CMB observed today.
This problem is aggravated in the cosmological context due to the fact 
that there were no observers in the early universe to carry out any 
measurements~\cite{qm-in-c}. 
One of the proposals which addresses the quantum measurement problem is the 
so-called Continuous Spontaneous Localization (CSL) model~\cite{csl}.
The advantage of using the CSL model to study the quantum measurement problem 
in the context of cosmology is that, in this model, the collapse of the 
wavefunction occurs without the presence of an observer.
In the CSL model, the Schr\"{o}dinger equation is modified by adding non-linear 
and stochastic terms which suppress the quantum effects in the classical 
domain, and also reproduce the predictions of quantum mechanics in the 
quantum regime (for reviews, see Refs.~\cite{csl-reviews}).  
In the context of inflation, there have been attempts to understand the 
quantum measurement problem by employing the CSL model~\cite{jerome-2012,
suratna-2013,Martin:2019jye,Martin:2019oqq,Bengochea:2020qsd,Martin:2020sdm,Bengochea:2020efe}.
Motivated by these efforts in the context of inflation, in this work, we 
shall investigate the quantum-to-classical transition in bouncing scenarios 
from the two perspectives described above.

\par

The remainder of this paper is organized as follows. 
In Sec.~\ref{sec:qsp}, working in the Schr\"{o}dinger picture, we shall quickly 
review the quantization of the tensor perturbations in an evolving universe 
and arrive at the wave function governing the perturbations. 
In Sec.~\ref{sec:tps-mb}, we shall review the evolution of the tensor perturbations 
in a specific classical matter bounce scenario and obtain the resulting 
tensor power spectrum.
This discussion will provide the background for appreciating
the effects of the CSL mechanism on the tensor power spectrum.
In Sec.~\ref{sec:stm}, using the Wigner function, we shall examine the squeezing
of the quantum state describing the tensor modes as they evolve in a matter bounce.
In Sec.~\ref{sec:i-csl-mb}, after a brief summary of the essential aspects of 
the CSL mechanism, we shall study its imprints on the tensor power spectrum 
produced in a matter bounce. 
In Sec.~\ref{sec:i-csl-tps-gb}, we shall discuss the evolution of the tensor
perturbations in a more generic classical bounce and evaluate the 
corresponding tensor power spectrum, including the effects due to CSL.
Finally, in Sec.~\ref{sec:csl-c}, we shall conclude with a brief summary of 
the main results.

\par

Note that we shall work with natural units wherein $\hbar=c=1$, and define 
the Planck mass to be $M_{_{\rm Pl}}=(8\, \pi\, G)^{-1/2}$. 
Working in $(3+1)$-spacetime dimensions, we shall adopt the metric signature 
of $(+,-,-,-)$.
Also, overprimes shall denote differentiation with respect to the conformal 
time coordinate~$\eta$.


\section{Quantization of the tensor perturbations in the Schr\"{o}dinger 
picture}\label{sec:qsp}
We shall consider a spatially flat Friedmann-Lema\^itre-Robertson-Walker (FLRW)
universe which is described by the line element 
\begin{equation}
\d s^2=a^2(\eta)\,\l(\d\eta^2-\delta_{ij}\,\d x^i\,\d x^j\r),
\end{equation}
where $a(\eta)$ denotes the scale factor, with $\eta$ being the conformal
time coordinate.
Upon taking into account the tensor perturbations, say, $h_{ij}$, the FLRW
metric assumes the form
\begin{equation}
\d s^2=a^2(\eta)\,\l[\d\eta^2-\l(\delta_{ij}+h_{ij}\r)\,\d x^i\,\d x^j\r],
\end{equation}
where $h_{ij}$ satisfies the traceless and transverse conditions  
(\ie~$h^i_i=0$ and~$\partial_jh^{ij}=0$).

\par

The second order action governing the tensor perturbations $h_{ij}$ is 
given by (see the following reviews~\cite{i-reviews})
\begin{equation}
\delta_2S=\frac{\Mpl^2}{8}\,
\int \d\eta\,\int\d^3 \x\; a^2(\eta)\, \l[h_{ij}'^2-(\partial h_{ij})^2\r].
\label{eq:ag-tp}
\end{equation}
The homogeneity and isotropy of the background metric permits the following
Fourier decomposition of the tensor perturbations:
\begin{equation}
h_{ij}(\eta,{\bm x})
=\sum_{s=1}^2\,\int \f{\d^3{\k}}{(2\,\pi)^{3/2}}\,
\varepsilon_{ij}^s(\k)\,
h_{\k}(\eta)\,{\rm e}^{i\,\k\cdot\x},
\end{equation}
where $\varepsilon_{ij}^s(\k)$ denotes the polarization tensor, with $s$ 
representing the helicity.
The polarization tensor satisfies the normalization 
condition:~$\varepsilon_{ij}^r(\k)\,
\varepsilon_{ij}^{s*}(\k)=2\,\delta^{rs}$~\cite{i-reviews}. 
In terms of the Fourier modes $h_{\k}$, the second order action~(\ref{eq:ag-tp}) 
can be expressed as
\begin{equation}
\delta_2S
=\frac{\Mpl^2}{2}\,\int \d\eta\, \int \d^3\k\; a^2(\eta)\,
\biggl[h_{\k}'(\eta)\, h_\k'^{*}(\eta)
- k^2\, h_{\k}(\eta)\,h_{\k}^{*}(\eta)\biggr],
\end{equation}
where $k =\vert\k\vert$. 
Note that, since $h_{ij}(\eta,{\bm x})$ is real, the integral over $\k$ runs 
over only half of the Fourier space, \ie~$\mathbb{R}^{3+}$.

\par

It proves to be convenient to express the tensor modes in terms of the 
so-called Mukhanov-Sasaki variable $u_\k$ as $h_{\k}= (\sqrt{2}/\Mpl)\,
(u_{\k}/a)$~\cite{i-reviews}.
In terms of the Mukhanov-Sasaki variable, the second order action~(\ref{eq:ag-tp}) 
takes the form
\begin{equation}
\delta_2S=\int \d \eta\, \int \d^3 {\k}\,\l[u_{{\k}}'\,u_{\k}'^{*}
-\omega_k^2(\eta)\,u_{\k}\, u_{\k}^{*}\r],\label{eq:ag-tp-msv}
\end{equation}
where 
\begin{equation}
\omega_k^2(\eta)= k^2-\f{a''}{a}.\label{eq:wk}
\end{equation}
It should be noted that, upon varying the action with respect to~$u_\k$,
one obtains the following equation of motion governing $u_\k$:
\begin{equation}
u_\k''+\omega_k^2(\eta)\,u_\k=0.\label{eq:uk-eom}
\end{equation}
The momenta associated with the variables $u_\k$ and $u_\k^*$ are given by
\begin{equation}
p_\k=u_\k'^*,\quad p_\k^*=u_\k'.
\end{equation}
The Hamiltonian associated with the above second order action can be determined 
to be
\begin{equation}
{\sf H}=\int \d^3 \k\,\l[p_{\k}\,p_{\k}^{*}+\omega_k^2(\eta)\,
u_{\k}\,u_{\k}^{*}\r].
\end{equation}
To carry out the quantization procedure, we need to deal with real variables 
(see, for instance, Refs.~\cite{dcp,jerome-2008}). 
Hence, let us write the variables $u_{\k}$ and $p_{\k}$ as 
\begin{equation}
u_{\k}= \f{1}{\sqrt{2}}\,\l(u_{\k}^{\R}+i\, u_{\k}^{\I}\r),\quad
p_{\k}= \f{1}{\sqrt{2}}\,\l(p_{\k}^{\R}+i\, p_{\k}^{\I}\r),
\end{equation}
where the superscripts $\R$ and $\I$ denote the real and imaginary parts 
of the corresponding quantities. 
In terms of these new variables, the Hamiltonian ${\sf H}$ is given by
\begin{equation}
{\sf H}=\int \d^3 {\k}\;{\sf H}_{\k}
=\int \d^3 \k\; \l({\sf H}_{\k}^{\R}+{\sf H}_{\k}^{\I}\r),
\end{equation}
where
\begin{equation}
{\sf H}_{{\k}}^{\R,\I}
=\f{1}{2}\,(p_{\k}^{\R,\I})^2
+\f{1}{2}\,\omega_k^2(\eta)\,(u_{\k}^{\R,\I})^2.\label{eq:Hk}
\end{equation}
It is evident from the structure of the Hamiltonian ${\sf H}$ that each 
variable $u_{\k}^{\R,\I}$ evolves independently as a parametric oscillator 
with the time-dependent frequency $\omega_k(\eta)$. 
Therefore, the complete quantum state of the system, say, $\Psi(u_\k,\eta)$, 
can be written as a product of the wavefunctions of the individual modes, 
say, $\psi_{\k}(u_{\k},\eta)$, in the following form: 
\begin{equation}
\Psi(u_\k,\eta)=\prod_{\k}\psi_{\k}(u_{\k},\eta)
=\prod_{\k}\psi_{\k}^{\R}(u_{\k}^{\R},\eta)\;
\psi_{\k}^{\I}(u_{\k}^{\I},\eta).
\end{equation}

\par

Quantization of the tensor perturbations can be achieved by promoting the 
variables $u_{\k}^{\R,\I}$ and $p_{\k}^{\R,\I}$ to quantum operators which
satisfy the following non-trivial canonical commutation relations:
\begin{equation}
\l[\hat u_{\k}^{\R},\,\hat p_{\k'}^{\R}\r]=i\,\delta^{(3)}(\k-\k'),\quad
\l[\hat u_{\k}^{\I},\hat p_{\k'}^{\I}\r]=i\,\delta^{(3)}(\k-\k').
\end{equation}
The Schr\"{o}dinger equation governing the evolution of the quantum state 
$\psi_{\k}^{\R,\I}$ corresponding to the mode $\k$ is given by
\begin{equation}
i\,\frac{\partial\psi_\k^{\R,\I} }{\partial \eta}
=\hat {\sf H}_{\k}^{\R,\I}\, \psi_{\k}^{\R,\I}.\label{eq:se} 
\end{equation}
Upon using the following representation for $\hat u_{\k}^{\R,\I}$ and 
$\hat p_{\k}^{\R,\I}$:
\begin{equation}
\hat u_{\k}^{\R,\I}\,\Psi=u_{\k}^{\R,\I}\,\Psi,\quad
\hat p_{\k}^{\R,\I}\,\Psi=-i\frac{\partial\Psi}{\partial u_{\k}^{\R,\I}},
\end{equation}
one can write the Hamiltonian operator in Fourier space 
$\hat {\sf H}_\k^{\R,\I}$ as 
\begin{equation}
\hat {\sf H}_{\k}^{\R,\I}
=-\f{1}{2}\,\frac{\pa^2}{\pa (u_{\k}^{\R,\I})^2}
+\f{1}{2}\,\omega_k^2(\eta)\,(\hat u_{\k}^{\R,\I})^2.\label{eq:Hko}
\end{equation}
It is well known that the wavefunction characterizing a time-dependent  
oscillator evolving from an initial ground state can be expressed as 
(see, for instance, Refs.~\cite{dcp})
\begin{equation}
\psi_{\k}^{\R,\I}(u_{\k}^{\R,\I},\eta)
=N_{k}(\eta)\,{\rm exp}-\l[\Omega_{k}(\eta)\,
(u_{\k}^{\R,\I})^2\r],\label{eq:wf}
\end{equation} 
where $N_k$ is the normalization constant which can be determined (up 
to a phase) to be $ N_{k}=\l(2\,\Omega_{k}^{\R}/\pi\r)^{1/4}$, with
$\Omega_{k}^{\R}$ denoting the real part of $\Omega_{k}$.
If we now write $\Omega_{k}=-(i/2)\,f_{k}'/f_{k}$ and substitute the above
wave function in the Schr\"{o}dinger equation~(\ref{eq:se}), then one finds that
the function $f_k$ satisfies the same classical equation of motion 
[\ie~Eq.~(\ref{eq:uk-eom})] as the Mukhanov-Sasaki variable~$u_\k$.
In other words, if we know the solution to the classical Mukhanov-Sasaki
equation, then we can arrive at the complete wavefunction $\psi_{\k}^{\R,\I}$
[cf. Eq.~(\ref{eq:wf})] describing the tensor modes. 
(Note that, since the equation governing $f_k$ and $u_\k$ are the same, 
hereafter, we shall often refer to $f_k$ as the tensor mode.)


\section{Tensor modes and power spectrum in a matter bounce}\label{sec:tps-mb}

We shall be interested in classical bouncing scenarios where the scale 
factor $a(\eta)$ is of the form 
\begin{equation}
a(\eta)=a_0\,\l[1+(\eta/\eta_0)^2\r]^q=a_0\,\l[1+(k_0\,\eta)^2\r]^q,
\label{eq:sf}
\end{equation}
where $a_0$ is the minimum value of the scale factor at the bounce (\ie~at 
$\eta=0$), $q$ is a positive real number, and ${\eta}_0={k_0}^{-1}$ is the 
time scale associated with the bounce. 
It is clear from the form of the scale factor that the universe starts in a
contracting phase at large negative $\eta$ with the scale factor reaching 
a minimum at $\eta = 0$, and expands thereafter.
Note that, at the bounce, since the scale factor attains a
minimum value, the Hubble parameter vanishes.
If we assume general relativity, in the spatially flat FLRW model of our 
interest, this implies that the total energy density should vanish at 
the bounce. 
Moreover, we require that $\dot{H}>0$ close to the bounce.
This implies that the sum of the energy density and pressure needs to be
negative around the bounce.
One finds that these behavior can be achieved with the aid of two fluids 
or two scalar fields such that their total energy density is zero at the 
bounce (in this context, see, for instance, 
Refs.~\cite{mb,rathul-2018,rathul-2019}).
For instance, if we consider a canonical scalar field driven by a suitable
potential such that its energy density behaves as matter and another 
non-canonical scalar field with only a kinetic term which acts as radiation,
but with negative energy density, then their dynamics will result in a bounce
with $q = 1$. 
A bounce with $q > 1$ can be achieved by modifying the parameters of the 
potential describing the canonical scalar field and changing the index of 
the kinetic term of the non-canonical scalar field.
Also, one finds that some of these models can lead to primordial scalar and 
tensor power spectra that are consistent with the CMB
data~\cite{rathul-2018,rathul-2019}.

\par

Before we discuss the case of evolution of the tensor modes in a 
classical bouncing universe characterized by an arbitrary 
value of $q$, it is instructive to consider the simpler case of $q = 1$.
Such a bounce is often referred to as a matter bounce, since, at early times, 
far away from the bounce, the scale factor behaves in the same manner as in 
a matter dominated era, \ie~as $a(\eta) \propto \eta^2$. 
The evolution of the tensor modes and the resulting power spectrum in such 
a matter bounce has been discussed before (see for instance, 
Refs.~\cite{starobinsky-1979,wands-1999,debika-2015}). 
For the sake of completeness, we shall briefly present the essential 
derivation here.

\par

We need to evolve the modes from early times during 
the contracting phase, across the bounce until a suitable time after 
the bounce, when we have to evaluate the power spectrum.
In order to arrive at an analytical expression for the tensor modes, 
it is convenient to divide this period of interest into two domains. 
Let the time range $-\infty<\eta<-\alpha\,\eta_0$ be the first domain, 
where the parameter $\alpha$ is a large number, say, $10^5$.
This period is far away from the bounce and corresponds to very early 
times, Since, in this domain, $\eta \ll -\eta_0$, the scale factor behaves 
as $a(\eta)\simeq a_0\,(k_0\,\eta)^2$. 
Therefore, the differential equation describing the tensor modes in the 
first domain reduces to
\begin{equation}
f_{k}''+\l(k^2-\f{2}{\eta^2}\r)f_{k}\simeq 0.\label{eq:ms-1}
\end{equation}
This is exactly the equation of motion satisfied by the tensor modes in 
de Sitter inflation, whose solutions are well known~\cite{wands-1999}.

\par

If we assume that, at very early times during the contracting phase, the 
oscillator corresponding to each tensor mode is in its ground state, then, 
we require that $\Omega_k=k/2$ for $\eta\ll-\eta_0$.
This, in turn, corresponds to demanding that, for $\eta\ll-\eta_0$, the 
tensor mode $f_k$ behaves as 
\begin{equation}
f_k(\eta) \simeq \f{1}{\sqrt{2\,k}}\,{\rm e}^{i\,k\,\eta},
\label{eq:bd-ic}
\end{equation}
which essentially corresponds to the Bunch-Davies initial condition, had we 
been working in the Heisenberg picture~\cite{bunch-1978}. 
Let $\eta_k$ be the time when $k^2=a''/a$, \ie~when the modes leave the Hubble 
radius during the contracting phase.
For cosmological modes such that $k/k_0\ll 1$, $\eta_k\simeq -\sqrt{2}/k$.
(If, say, $k_0/a_0\simeq \Mpl$, one finds that $k/k_0$ is of the order of 
$10^{-28}$ or so for scales of cosmological interest.)
The Bunch-Davies initial condition can be imposed when $\eta\ll\eta_k$.
We shall assume that $\eta_k\ll-\alpha\,\eta_0$, which
corresponds to $k\ll k_0/\alpha$.
Since, as we mentioned, Eq.~(\ref{eq:ms-1}) resembles that of the equation in 
de Sitter inflation, the tensor mode $f_k$ satisfying the Bunch-Davies initial 
condition can be immediately written down to be~\cite{starobinsky-1979,
wands-1999,debika-2015}
\begin{equation}
f_{k}^{({\rm I})}(\eta)=\f{1}{\sqrt{2\,k}}\,
\l(1+\f{i}{k\,\eta}\r)\,{\rm e}^{i\,k\,\eta}.
\end{equation}

\par

The solution $f_k^{({\rm I})}$ we have obtained above is valid
in the first domain, \ie~over $-\infty<\eta<-\alpha\,\eta_0$. 
Let us now turn to the evolution of the mode during the second domain, 
which covers the period of bounce.
The domain corresponds to $-\alpha\,\eta_0<\eta<\beta\,\eta_0$, where we 
shall assume $\beta$ to be of the order of $10^2$.
Over this domain, for scales of our interest (\ie~$k\ll k_0/\alpha$), we 
can ignore the $k^2$ term in Eq.~(\ref{eq:uk-eom}) which governs $f_k$.
In such a case, the equation simplifies to
\begin{equation}
f_k''-\f{a''}{a}f_{k}\simeq 0\label{eq:ms-2}
\end{equation}
or, equivalently, in terms of the original variable $h_k$, to
\begin{equation}
h_{k}''+2\,\f{a'}{a}\,h_{k}'\simeq 0.
\end{equation}
Using the exact form~(\ref{eq:sf}) of the scale factor, this equation 
can be immediately integrated to obtain the following solution in the 
second domain~\cite{debika-2015}:
\begin{equation}
f_k^{({\rm II})}(\eta)=a(\eta)\,\l[A_k+B_k\,g(k_0\,\eta)\r],
\end{equation}
where $A_k$ and $B_k$ are constants, and the function $g(x)$ is given by 
\begin{equation}
g(x) = \f{x}{1+x^2}+{\rm tan}^{-1}(x).\label{eq:g}
\end{equation}
The constants $A_k$ and $B_k$ are arrived at by matching the solutions 
$f_k^{({\rm I})}$ and $f_k^{({\rm II})}$ and their derivatives with
respect to $\eta$ at $-\alpha\,\eta_0$.
We find that $A_k$ and $B_k$ are given by
\begin{subequations}
\begin{eqnarray}
A_k &=& \frac{1}{\sqrt{2\,k}}\,\l(\f{1}{a_0\,\alpha^2}\r)\,
\l(1-\f{i\,k_0}{\alpha\, k}\r)\,{\rm e}^{-i\,\alpha\, k/k_0}
+B_k\, g(\alpha),\nn\\
\\
B_{k} &=&\f{1}{\sqrt{2\,k}}\,
\f{\l(1+\alpha^2\r)^2}{2\,a_0\,\alpha ^2}\,
\l(\f{i\,k}{k_0}+\f{3}{\alpha}-\f{3\,i\,k_0}{\alpha^2\, k}\r)\,
{\rm e}^{-i\,\alpha\, k/k_0}.\nn\\
\end{eqnarray}
\end{subequations}

\par

Assuming that the universe transits to the conventional
radiation dominated epoch at the end of the second domain, we evaluate 
the tensor power spectrum at $\eta=\beta\,\eta_0$ after the 
bounce~\cite{debika-2015}.
Recall that the tensor power spectrum is defined in terms of the mode 
function $f_k(\eta)$ as~\cite{i-reviews}
\begin{equation}
{\cal P}_{_{\rm T}}(k)
=\f{8}{\Mpl^2}\,\f{k^3}{2\,\pi^2}\,
\f{\vert f_k(\eta)\vert^2}{a^2(\eta)}.\label{eq:tps-d}
\end{equation}
On using the solution $f_k^{({\rm II})}$ above, the tensor power spectrum at 
$\eta=\beta/k_0$ can be expressed as
\begin{equation}
{\cal P}_{_{\rm T}}(k)
=\f{8}{\Mpl^2}\,\f{k^3}{2\,\pi^2}\,
\vert A_k+B_k\,g(\beta)\vert^2.
\end{equation}
Note that our analytical expressions and the resulting power spectrum are valid
only for modes such that $k\ll (k_0/\alpha)$. 
Over such a range of wavenumbers, for a large enough $\beta$, it is 
straightforward to show that the tensor power spectrum is strictly 
scale invariant and has the amplitude~\cite{rathul-2018,debika-2015} 
\begin{equation}
{\cal P}_{_{\rm T}}(k)\simeq \frac{9\,k_0^2}{2\,a_0^2\,\Mpl^2}.
\end{equation}
Such a scale invariant spectrum is indeed expected to arise in a matter bounce
as the scenario is `dual' to de Sitter inflation~(in this context, see
Ref.~\cite{wands-1999}).


\section{Squeezing of quantum states associated with tensor modes in the matter bounce}\label{sec:stm}

Having discussed the evolution of the tensor modes through a 
classical matter bounce, 
let us turn our attention to the behavior of the quantum state~$\psi_{\k}$. 
We shall essentially follow the approach adopted in the context of perturbations 
generated during inflation~\cite{dcp,squeezing,jerome-2012,suratna-2013}.

\par

In classical mechanics, one of the ways of understanding the evolution of a 
system is to examine its behavior in phase space. 
However, since canonically conjugate variables cannot be measured simultaneously
in quantum mechanics, a method needs to be devised in order to compare the 
evolution of a quantum system with its classical behavior in phase space.
As is well known, one of the ways to understand the evolution of a quantum 
state is to examine the behavior of the so-called Wigner function, which is 
a quasi-probability distribution in phase space that can be constructed 
from a given wave function. 
Recall that the wave function corresponding to a tensor mode can be expressed
as
\begin{equation}
\psi_{\k}(u_\k,\eta)
=\psi_{\k}^\R(u_\k^\R,\eta)\;\psi_{\k}^\I(u_{\k}^\I,\eta)
=N_{k}^2\,{\rm exp}-\l(2\,\Omega_{k}\,u_\k\,u_\k^*\r),\label{eq:cwf}
\end{equation}
where, as mentioned before, $N_{k}=(2\,\Omega_{k}^{\R}/\pi)^{1/4}$, $\Omega_{k}
=-(i/2)\,f_{k}'/f_{k}$ 
and $f_k$ satisfies the differential equation~(\ref{eq:uk-eom}). 
The Wigner function associated with the quantum state~(\ref{eq:cwf}) 
is defined as~\cite{dcp,jerome-2012,suratna-2013}
\begin{eqnarray}
W(u_{\k}^{\rm R},u_{\k}^{\rm I},p_{\k}^{\rm R},p_{\k}^{\rm I},\eta)
&=& \f{1}{(2\,\pi)^2}\,
\int_{-\infty}^{\infty}\d x\;\int_{-\infty}^{\infty}\d y\;
\psi_{\k}\l(u_{\k}^{\rm R}+\f{x}{2},u_{\k}^{\rm I}+\f{y}{2},\eta\r)\nn\\
& &
\times\,\psi^*_{\k}\l(u_{\k}^{\R}-\f{x}{2},u_{\k}^{\rm I}-\f{y}{2},\eta\r)\, 
{\rm exp} -i\,\l(p_{\k}^{\rm R}\, x+p_{\k}^{\rm I}\, y\r).
\end{eqnarray}
The integrals over $x$ and $y$ can be easily evaluated to arrive at 
the following form for the Wigner function~\cite{jerome-2012} 
\begin{eqnarray}
W(u_{\k}^{\R},u_{\k}^{\I},p_{\k}^{\R},p_{\k}^{\I},\eta)
&=&\f{\l\vert\psi_{\k}(u_{\k},\eta)\r\vert^2}{2\,\pi\,\Omega_k^\R}\; 
{\rm exp}-\l[\f{1}{2\,\Omega_{k}^\R}\,
\l(p_{\k}^{\rm R}+2\,\Omega_{\k}^\I\,u_{\k}^{\rm R}\r)^2\r]\nn\\
&&\times\, {\rm exp}-\l[\f{1}{2\,\Omega_{k}^\R}\,
\l(p_{\k}^{\rm I}+2\,\Omega_{\k}^\I u_{\k}^{\rm I}\r)^2\r].
\end{eqnarray}

\par

\begin{figure}[!t]
\begin{center}
\begin{tabbing}
\includegraphics[width=0.45\textwidth]{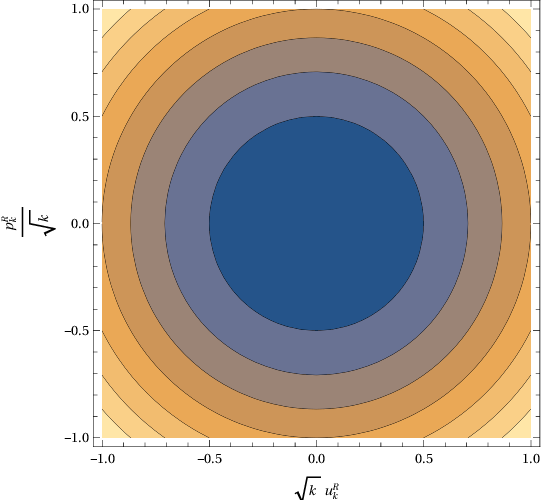}
\=\hspace{8pt}\includegraphics[width=0.45\textwidth]{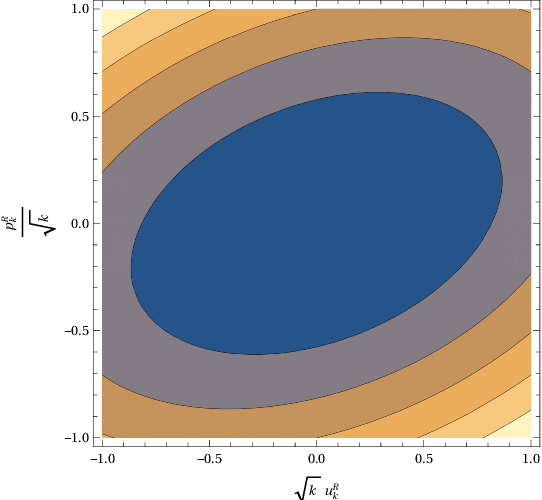}\\
\vspace{8pt}\\
\includegraphics[width=0.45\textwidth]{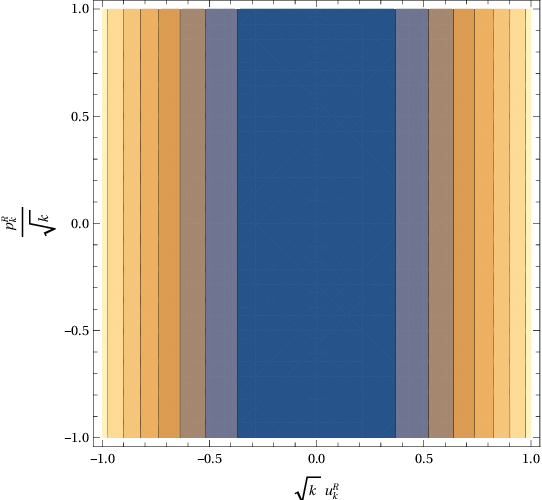}\>
\=\hspace{8pt}
\includegraphics[width=0.45\textwidth]{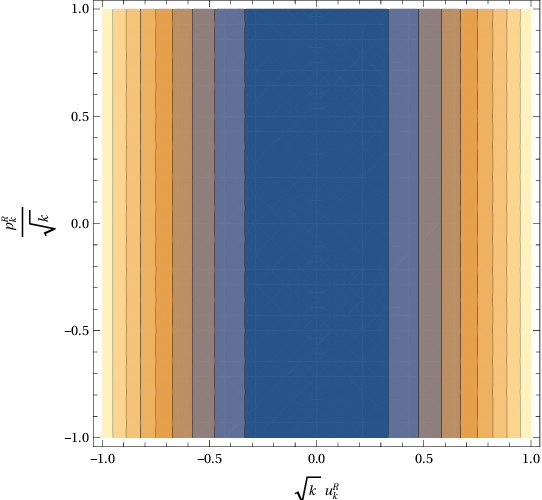}\\
\end{tabbing}
\caption{\label{fig:Wigner}The evolution of the Wigner function 
$W(u_{\k}^{\rm R},u_{\k}^{\rm I},p_{\k}^{\rm R},p_{\k}^{\rm I},\eta)$ associated 
with the quantum state that describes a tensor mode of cosmological interest.
Out of the two independent sets of variables $(u_{\k}^{\rm R},p_{\k}^{\rm R})$ 
and $(u_{\k}^{\rm I},p_{\k}^{\rm I})$, we have chosen the set $(u_{\k}^{\rm R},
p_{\k}^{\rm R})$ and have fixed $(u_{\k}^{\rm I},p_{\k}^{\rm I})=(0,0)$ to 
illustrate the behavior of the quantity $\ln\, [W(u_{\k}^{\rm R},u_{\k}^{\rm I},
p_{\k}^{\rm R},p_{\k}^{\rm I},\eta)]$.
In plotting these figures, we have set $k_0/(a_0\,\Mpl)=10^{-5}$, and have chosen the mode corresponding to $k/k_0=10^{-15}$.
The plots correspond to the times ${\cal N}=-12.51$ (top left), ${\cal N} = -11.75$ 
(top right), ${\cal N}=0$ (bottom left) and  ${\cal N} = 4.29$ (bottom right).
The first two instances (\viz~when ${\cal N}=-12.51$ and ${\cal N} = -11.75$)
correspond to situations when the mode is in the strongly sub-Hubble domain
and close to Hubble exit during the contracting phase, respectively.
Note that, as time evolves, the Gaussian state that is initially symmetric 
in $u_{\k}^{\rm R}$ and $p_{\k}^{\rm R}$ (top left) gets increasingly squeezed 
about about $u_\k^{\rm R}=0$ (top right, bottom left) as one approaches the 
bounce, and remains so (bottom right) as the universe begins to expand.
This largely reflects the behavior that occurs in the inflationary 
scenario.}\label{fig:wf} 
\end{center}
\end{figure}

Since we know the mode functions $f_k$, we can evaluate $\Omega_k^\R$
and $\Omega_k^\I$ and thereby determine the above Wigner function as a
function of time.
Note that, in inflation, to cover a wide range in time, one often works 
with e-folds, say, $N$, as the time variable.
The e-folds are defined through the relation $a(N)=a_{\rm i}\, {\rm exp}\,
(N-N_{\rm i})$, where, evidently, $a=a_{\rm i}$ at $N=N_{\rm i}$.
However, the exponential function ${\rm e}^{N}$ is a monotonically growing 
function and hence does not seem appropriate to describe bounces.
In the context of bounces, particularly the symmetric ones of our 
interest, it seems more suitable to introduce a new variable $\cN$ 
known as e-N-folds, which is defined through the 
relation $a(\cN)=a_0\,{\rm exp}\l({\cal N}^2/2\r)$~\cite{sriram-2015}. 
In the matter bounce, the conformal time coordinate $\eta$ is related to
e-N-folds as
\begin{equation}
\eta({\cal N})=\pm k_0^{-1}\l(e^{{\cal N}^2/2}-1\r)^{1/2},
\end{equation}
with $\cN$ being zero at the bounce, while it is negative before the 
bounce and positive after. 
Using the above relation $\eta({\cal N})$, we have converted the Wigner
function as a function of $\cN$.
In Fig.~\ref{fig:wf}, we have illustrated the behavior of the function in
terms of contour plots in the $(u_{\k}^{\rm R},p_{\k}^{\rm R})$-plane as
a tensor mode (corresponding to a scale of cosmological interest) evolves 
across the bounce. 

\par

\begin{figure}[!t]
\begin{center}
\includegraphics[width=8.50cm]{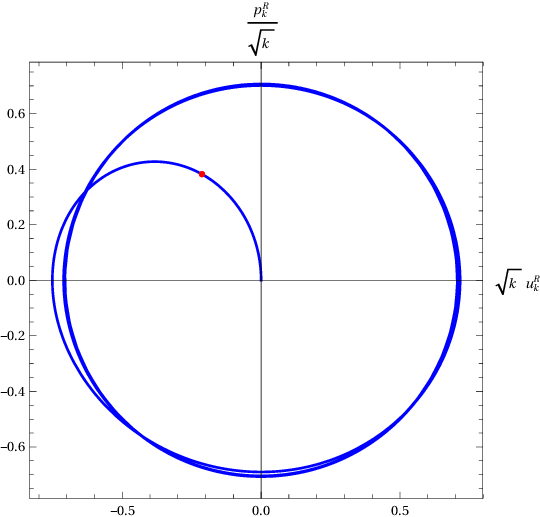} 
\caption{\label{fig:pp}The behavior of a typical cosmological mode in the 
dimensionless classical $(\sqrt{k}\,u_\k^\R,p_\k^\R/\sqrt{k})$-phase plane.
We have chosen a mode such that $k/k_0=10^{-15}$.
The red dot indicates the time at which the mode leaves the Hubble radius 
 during the contracting phase.}
\end{center}
\end{figure}

It is easy to understand the above behavior of the Wigner function and, 
in particular, argue that the function peaks on the classical trajectory 
at late times.
Since $f_k$ satisfies the same equation as $u_\k$, at early times such 
that $k\,\eta\to -\infty$, we have
\begin{equation}
\sqrt{k}\,u_\k^\R(\eta)\simeq \f{1}{\sqrt{2}}\,{\rm cos}\,(k\,\eta),\quad
\f{p_\k^\R(\eta)}{\sqrt{k}}\simeq -\f{1}{\sqrt{2}}\,{\rm sin}\,(k\,\eta),
\end{equation}
which describes a circle of radius half in the $(\sqrt{k}\,u_\k^\R,
p_\k^\R/\sqrt{k})$-plane.
Let us consider a mode such that it reaches super-Hubble scales in the first
domain during the contracting phase, \ie~$-\sqrt{2}/k\ll-\alpha\,\eta_0$ 
or, equivalently, $k/(k_0/\alpha)\ll 1$.
This is exactly the limit under which we have constructed the analytical 
solutions earlier.
Therefore, on super-Hubble scales, we obtain that
\begin{equation}
\sqrt{k}\,u_\k^\R(\eta)\simeq \f{-1}{\sqrt{2}}\, \f{(k\,\eta)^2}{3},\quad
\f{p_\k^\R(\eta)}{\sqrt{k}}\simeq \f{-1}{\sqrt{2}}\, \f{2\,(k\,\eta)}{3},
\end{equation}
which implies that, in this limit, we have 
\begin{equation}
\f{p_\k^\R(\eta)}{2\sqrt{k}}\simeq \f{\sqrt{k}\,u_\k^\R(\eta)}{k\,\eta}.
\end{equation}
In other words, on super-Hubble scales during the contracting phase, both 
$\sqrt{k}\,u_\k^\R$ and $p_\k^\R/\sqrt{k}$ approach zero with the coordinate 
$\sqrt{k}\,u_\k^\R$ approaching zero faster than the conjugate 
momentum $p_\k^\R/\sqrt{k}$.
One finds that this behavior continues in the second domain and after the
bounce. 
In Fig.~\ref{fig:pp}, using the analytical solutions, we have plotted the 
behavior of a typical cosmological mode in the $(\sqrt{k}\,u_\k^\R,
p_\k^\R/\sqrt{k})$-plane.
The behavior we have described above is clearly illustrated by the phase space
trajectory in the figure.

\color{black}

\par

In a time-dependent background, the modes associated with quantum 
fields are generally expected to get increasingly squeezed as time 
evolves~\cite{squeezing}.
Let us now try to understand the extent to which the tensor modes
are squeezed in the matter bounce scenario.
If we define $f_k$ as~\cite{jerome-2012}
\begin{equation}
f_{k} = \f{1}{\sqrt{2\,k}}\,(\tilde u_{k}+{\tilde v_{k}^*}), 
\end{equation}
then the second order differential equation~(\ref{eq:uk-eom}) governing
$f_k$ can be written as two coupled first order differential equations 
as follows:
\begin{equation}
\tilde u_{k}'=i\,k\,\tilde u_{k}+\f{a'}{a}\,\tilde v_{k}^*,\quad
\tilde v_{k}'=i\,k\,\tilde v_{k}+\f{a'}{a}\,\tilde u_{k}^*.
\label{eq:ukt-vkt}
\end{equation}
The Wronskian, say, ${\sf W}$, corresponding to the equation governing $f_k$ 
is defined as ${\sf W}= f_{k}'\,f_{k}^*-f_{k}'^{*}\,f_{k}$.
It can be readily shown using equation~(\ref{eq:uk-eom}) that $\d {\sf W}/\d\eta
=0$ or, equivalently, ${\sf W}$ is a constant.
If we assume that the modes $f_k$ satisfy the Bunch-Davies initial 
condition~(\ref{eq:bd-ic}), then one finds that ${\sf W}=i$.

\par

In terms of $\tilde{u}_{k}$ and $\tilde{v}_{k}$, the Wronskian can be expressed
as ${\sf W}=i\,(\vert\tilde u_{k}\vert^2-\vert\tilde v_{k}\vert^2)$. 
Since ${\sf W}=i$, we can parametrize the variables $\tilde{u}_{k}$ and 
$\tilde{v}_{k}$ as~\cite{jerome-2008,jerome-2012} 
\begin{equation}
\tilde{u}_k={\rm e}^{i\,\theta_{k}}\,{\rm cosh}\,(r_{k}),\quad
\tilde{v}_k={\rm e}^{-i\,\theta_{k}+2\,i\,\phi_{\k}}\,{\rm sinh}\,(r_{k}),
\label{eq:rk-d}
\end{equation}
where $r_k$, $\theta_k$ and $\phi_k$ are known as the squeezing 
parameter, the rotation and squeezing angles, respectively. 
On substituting the expressions~(\ref{eq:rk-d}) in Eqs.~(\ref{eq:ukt-vkt}), 
one can arrive at a set of coupled differential equations which determine 
the behavior of the parameters $r_k$, $\theta_k$ and $\phi_k$ with respect 
to $\eta$ \cite{jerome-2012,jerome-2008}. 
The coupled differential equations governing these parameters are given by 
\begin{subequations}
\label{eq:s-eom}
\begin{eqnarray}
r_{k}'&=&\f{a'}{a}\,{\rm cos}\,(2\,\phi_{k}),\label{eq:rk-eom}\\
\phi_{k}'&=&k-\f{a'}{a}\,{\rm coth}\,(2\,r_{k})\,{\rm sin}\,(2\,\phi_{k}),\\
\theta_{k}'&=&k-\f{a'}{a}\,{\rm tanh}\,(r_{k})\,{\rm sin}\,(2\,\phi_{k}).
\end{eqnarray}
\end{subequations}
Our primary quantity of interest is the parameter $r_{k}$ which
characterizes the extent of squeezing of the quantum state 
$\psi_{\k}(u_{\k},\eta)$ as the universe evolves~\cite{squeezing}.

\par

\begin{figure}[!t]
\begin{center}
\includegraphics[width=8.50cm]{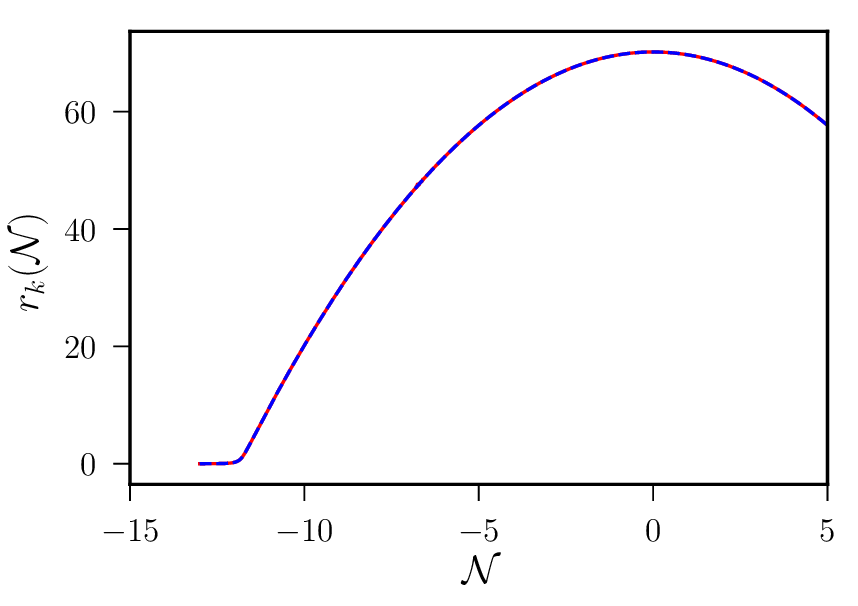} 
\caption{The analytical (in red) and the numerical (in blue) solutions 
for the squeezing parameter $r_k$ have been plotted as a function of 
e-N-folds, for a mode corresponding to $k/k_0=10^{-15}$ and values
of the parameters of the model mentioned in the earlier figures. 
Since we begin with the Bunch-Davies initial condition at very early times, 
the squeezing parameter $r_k$ is close to zero.
As the universe contracts, $r_k$ increases till it reaches a maximum at 
the bounce.
Then it decreases to some extent after the bounce, before the universe is 
assumed to enter the radiation dominated era.}\label{fig:s}
\end{center}
\end{figure}

By assuming the scale factor of interest, one can attempt to solve the 
differential equations~(\ref{eq:s-eom}) to arrive at the behavior of 
the squeezing parameter.
These equations essentially stem from the original equation~(\ref{eq:uk-eom})
that determines the evolution of the Mukhanov-Sasaki variable $u_k$ or $f_k$.
Since, we already know the solution to $f_k$ across the bounce, it would 
be simpler to express the parameters $r_k$, $\theta_k$ and $\phi_k$ in 
terms of $f_k$.
To begin with, we find that the variables $\tilde u_{k}$ and $\tilde v_{k}$ 
can be expressed in terms of $f_k$ and its derivative $f_k'$ as follows:
\begin{subequations}
\begin{eqnarray}
\tilde u_{k}&=&\sqrt{\frac{k}{2}}\,\l(1+\f{i}{k}\,\f{a'}{a}\r)\,f_k
-\f{i}{\sqrt{2\,k}}\,f_k',\\
\tilde v_{k}
&=&\sqrt{\frac{k}{2}}\,\l(1+\f{i}{k}\,\f{a'}{a}\r)\,f_k^\ast
-\f{i}{\sqrt{2\,k}}\,f_k'^\ast,
\end{eqnarray}
\end{subequations}
and it is straightforward to examine that $\vert\tilde u_{k}\vert^2
-\vert\tilde v_{k}\vert^2=1$, as required.
Once we have these two quantities at hand, we can obtain the squeezing 
parameters $r_k$, $\phi_k$ and $\theta_k$ from the relations
\begin{equation}
r_{k}={\rm sinh}^{-1}\l(\vert\tilde v_{k}\vert\r),\;\;
\phi_k=\f{1}{2}\,{\rm Arg}\,(\tilde u_{k} \tilde v_{k}),\;\;
\theta_k={\rm Arg}\,(\tilde u_{k}).
\end{equation}
Using the solutions for $f_k$ we have obtained in the case of the 
matter bounce in the previous section, we have plotted the behavior 
of the squeezing parameter $r_k$ as a function of e-N-folds 
in Fig.~\ref{fig:s}.
We have also independently solved the differential equations~(\ref{eq:s-eom})
using {\sl Mathematica}\/~\cite{Mathematica} to check the validity of the
analytical solution for $r_k$.
The numerical solution has also been plotted in the figure.
The agreement between the solutions clearly indicate the extent of
accuracy of the analytical solutions.
It is evident from the figure that the wave function associated with
the quantum state $\psi_\k$ is increasingly squeezed as the universe
evolves, reaching a maximum at the bounce.
In fact, it is this behavior which was reflected in the behavior
of the Wigner function (which had peaks about $u_\k^{\rm R}=0$) 
we had considered earlier.
While there are similarities in the behavior with what occurs in inflation,
there are some crucial differences as well.
In inflation, the parameter increases indefinitely with the duration of 
inflation~\cite{jerome-2012}.
For a duration corresponding to about $60$ e-folds of inflation, as is
typically required to overcome the horizon problem, $r_k$ is found to
grow to about $10^2$ or so.
We find that $r_k$ grows to the same order of magnitude in the matter 
bounce as well. 
However, we should clarify that the exact extent of 
the growth of $r_k$ will depend on when the modes leave the Hubble 
radius during the contracting phase prior to the bounce.
In accordance with the Heisenberg's uncertainty principle, squeezing 
of the quantum state about $u_{\k}^{\rm R}=0$ gives infinite 
possibilities of the momentum variable $p_{\k}^{\rm R}$. 
Hence, a squeezed state is not strictly a classical state. 
But, it has been argued that in a strongly squeezed quantum 
state, the vacuum expectation values and the stochastic mean 
are indistinguishable, if the perturbations are assumed to be 
realizations of a classical stochastic process~\cite{dcp}.
In fact, this comment can be expected to be true
for fluctuations about the mean as well.
In such a sense, one can argue that in the extreme squeezed limit 
the quantum state `turns' classical. 
However, in contrast to inflation where the growth seems indefinite, 
in the matter bounce, the parameter~$r_k$ begins to decrease as the 
universe begins to expand.
This interesting behavior may point to crucial differences between the 
quantum-to-classical transition in inflation and bounces and seem to
require further study.


\section{CSL modified tensor modes and power spectrum in the matter bounce}\label{sec:i-csl-mb}

As we had described in the introduction, one can also view the transition 
of primordial quantum perturbations into the classical LSS as a quantum 
measurement problem. 
In other words, we need to understand as to how the mechanism by which the 
original state of the primordial perturbations collapsed into a particular 
eigenstate which corresponds to the realization of the CMB observed today.
One of the proposals which addresses this issue is known as the CSL 
model~\cite{csl}. 
The crucial advantage of this model is that a specific realization can be 
attained without the presence of an observer.
In the rest of the manuscript we will focus on understanding the effects 
of CSL on the tensor perturbations in bouncing universes. 


\subsection{CSL in brief}

The CSL model proposes a unified dynamical description which suppresses the 
quantum effects, such as the superposition of states in the macroscopic regime, 
and reproduces the predictions of quantum mechanics in the microscopic regime. 
In CSL, a unified dynamical description is achieved by appropriately modifying
the Schr\"{o}dinger equation. 
This modification is carried out by adding nonlinear terms and a stochastic 
behavior which is encoded through a Wiener process~\cite{csl}. 
The modified Schr\"{o}dinger equation encompasses an amplification mechanism 
which makes the new terms negligible in the quantum regime, hence retrieving 
the dynamics predicted by quantum mechanics. 
At the same time, it should make the new terms dominant in the classical regime, 
so that a suitable behavior of the system is attained in the classical 
domain (for reviews, see Refs.~\cite{csl-reviews}). 
Although, it should be clarified that, in the implementation of CSL for 
the case of primordial perturbations~\cite{jerome-2012,suratna-2013}, 
the above mentioned amplification mechanism does not arise.

\par

A fully relativistic implementation of CSL does not yet 
exist (see Refs.~\cite{Tumulka:2005ki,Bedingham:2010hz,Bedingham:2011}). 
Hence, certain assumptions are required to be made. 
Depending on the assumptions that one makes, there are different versions 
of CSL (in this regard, see the discussion in Ref.~\cite{Martin:2019oqq}).
In this work, we follow the version of CSL implemented originally in the 
context of inflation~\cite{jerome-2012,suratna-2013}. 
In this version of CSL, one works in Fourier space and assumes that the 
primordial perturbations are Gaussian. 
Such an assumption seems justified since the non-Gaussianities associated
with the scalar perturbations are strongly constrained by the Planck
satellite~\cite{Akrami:2019izv}. 
(Of course, the primordial tensor perturbations remain to be detected.
However, both in inflation and bounces, the non-Gaussianities associated
with the tensors seem to be small~\cite{debika-2015}.)
One of the key assumptions in implementing CSL is the choice of the 
collapse operator. 
We have chosen a situation wherein the collapse operator depends on 
the Mukhanov-Sasaki variable.
Upon taking into account such effects, the modified Schr\"{o}dinger 
equation is given by~\cite{jerome-2012,suratna-2013} 
\begin{equation}
\d\psi_\k^{\R,\I}
=\l[-i\,\hat{\sf H}_{\k}^{\R,\I}\,\d\eta
+\sqrt{\gamma}\,\l(\hat u_{\k}^{\R,\I}-{\bar u}_{\k}^{\R,\I}\r)\,\d {\cal W}_{\eta}
-\f{\gamma}{2}\,\l(\hat u_{\k}^{\R,\I}-{\bar u}_{\k}^{\R,\I}\r)^2\,\d \eta\r]\,
\psi_{\k}^{\R,\I},\label{eq:mse}
\end{equation}
where $\hat{\sf H}_{\k}$ is the original Hamiltonian operator~(\ref{eq:Hko}),
$\gamma$ is the CSL parameter, which is a measure of the strength 
of the collapse and ${\cal W}_\eta$ denotes a real Wiener process, which 
is responsible for the stochastic behavior. 
If the CSL modified wavefunction $\psi_{\k}^{\R,\I}$ is assumed to be of 
the following form~\cite{bassi-2005,jerome-2012,suratna-2013} 
\begin{equation}
\psi_{\k}^{\R,\I}(u_{\k}^{\R,I},\eta)
=N_{k}(\eta)\, \ {\rm exp}-\l[\Omega_{k}(\eta)\,
\l(u_{\k}^{\R,\I}-{\bar u}_{\k}^{\R,\I}\r)^2
+i\,\chi_{\k}^{\R,\I}(\eta)\,u_{\k}^{\R,\I}
+i\,\sigma_{\k}^{\R,\I}(\eta)\r],
\end{equation}
then the functions $\Omega_{k}(\eta)$, $\chi_{\k}(\eta)$ and $\sigma_{\k}(\eta)$ 
are found to satisfy the following set of differential equations: 
\begin{subequations}
\begin{eqnarray}
\Omega_{k}'&=&-2\,i\,\Omega_{k}^2+\f{i}{2}\,\omega_k^2+\f{\gamma}{2},\
\label{eq:m-mse}\\
\f{N_{k}'}{N_{k}}&=&\Omega_{k}^\I,\\
\l({{\bar u}_{\k}^{\R,\I}}\r)'
&=&\chi_{\k}^{\R,\I}
+\f{\sqrt{\gamma}}{2\,\Omega_{k}^\R}\;{{\cal W}_{\eta}}',\\ 
\l(\chi_{\k}^{\R,\I}\r)'
&=&-\omega_k^2\, \bar{u}_{\k}^{\R,\I}
-\sqrt{\gamma}\,\f{\Omega_{k}^\I}{\Omega_{k}^\R}\,{{\cal W}_{\eta}}',\\
\l(\sigma_{\k}^{\R,\I}\r)'
&=&\f{\omega_k^2}{2}\,\l(\bar{u}_{\k}^{\R,\I}\r)^2
-\f{1}{2}\,\l(\chi_{\k}^{\R,\I}\r)^2-\Omega_{k}^\R
+\sqrt{\gamma}\,\f{\Omega_{k}^\I}{\Omega_{k}^\R}\,
\bar{u}_{\k}^{\R,\I}\,{{\cal W}_{\eta}}',
\end{eqnarray}
\end{subequations}
where $\omega_k^2$ is given by Eq.~(\ref{eq:wk}).

\par

In principle, one needs to solve the above set of stochastic differential
equations in order to arrive at a complete understanding of the effects of 
CSL.
However, recall that, our primary concern is the imprints of CSL on the 
tensor power spectrum.
Note that, earlier, we had defined $\Omega_k=-(i/2)\,f_k'/f_k$ and the 
original Schr\"{o}dinger equation had led to $f_k$ satisfying the
Mukhanov-Sasaki equation~(\ref{eq:uk-eom}).
If we now substitute the same expression for $\Omega_k$ in the CSL
corresponding modified equation~(\ref{eq:m-mse}), we find that $f_k$ 
now satisfies the differential equation~\cite{jerome-2012}
\begin{equation}
f_k''+\l(k^2-i\,\gamma-\f{a''}{a}\r)f_k=0,\label{eq:m-uk-eom}
\end{equation}
\ie~the effects of CSL is essentially to replace $k^2$ by $k^2-i\,\gamma$.
We should clarify that the above modification to
the Mukhanov-Sasaki equation is valid in any spatially flat FLRW
universe.
Hence, just as it has been utilized to examine the effects of CSL
in the inflationary scenario, it can be used equally well in the
case of the bouncing scenario of our interest here.
In the following sub-section, we shall solve this equation in a matter
bounce and evaluate the effects of CSL on the tensor power spectrum. 


\subsection{CSL modified tensor power spectrum}\label{eq:e-csl-tps}

In this sub-section we shall focus on the evaluation of CSL modified 
tensor power spectrum in the classical matter bounce scenario.

\par

We find that, under certain conditions, even the CSL modified modes can 
be arrived at using the approximations we had worked with earlier.
If we divide the period of our interest into two domains, we find that
the CSL modified modes in the first domain (\ie~over $-\infty<\eta<-\alpha\,
\eta_0$), which satisfy the Bunch-Davies initial conditions, can be 
expressed as (for a discussion on the initial conditions for the case of 
CSL modified tensor modes, see Ref.~\cite{jerome-2012})
\begin{equation}
f_{k}^{(\rm I)}(\eta)
=\f{1}{\sqrt{2\, z_k\, k}}\l(1+\f{i}{z_k\, k\,\eta}\r)\,
{\rm e}^{i\,z_k\, k\,\eta},\label{eq:mfk-1}
\end{equation}
where $z_k=\l(1-i\,\gamma/k^2\r)^{1/2}$.

\par

In the second domain, \ie~in the time range $-\alpha\,\eta_0<\eta<\beta\,
\eta_0$, the term $a''/a$ in Eq.~(\ref{eq:m-uk-eom}) behaves as $a''/a
\geq 2\,k_0^2/(1+\alpha^2)$. 
Recall that the modes of cosmological interest are assumed to be very small 
compared to $k_0/\alpha$.
Hence, if the CSL parameter $\gamma$ is also assumed to be very small 
when compared to $k_0^2$, then Eq.~(\ref{eq:m-uk-eom}) can be approximated 
to be
\begin{equation}
f_k''-\f{a''}{a}\,f_k\simeq 0,
\end{equation}
exactly as in the unmodified case.
Upon integrating this equation, we obtain that 
\begin{equation}
f_k^{({\rm II})}(\eta)
=a(\eta)\,\l[A_k^{(\gamma)}+B_k^{(\gamma)}\,g(k_0\eta)\r],
\label{eq:mfk-2}
\end{equation}
where $g(x)$ is the same function~(\ref{eq:g}) we had encountered earlier, 
while $A_k^{(\gamma)}$ and $B_k^{(\gamma)}$ are given by 
\begin{subequations}
\begin{eqnarray}
A_k^{(\gamma)} 
&=& \frac{1}{\sqrt{2\,z_k\, k}}\,
\f{1}{a_0\,\alpha^2}\,\l(1-\f{i\,k_0}{\alpha\, z_k\, k}\r)\,
{\rm e}^{-i\,\alpha\, z_k\, k/k_0}
+\,B_k^{(\gamma)}\, g(\alpha),\\
B_{k}^{(\gamma)}
&=&\frac{1}{\sqrt{2\,z_k\, k}}\,
\f{\l(1+\alpha^2\r)^2}{2\,a_0\,\alpha^2}\,
\l(\f{i\,z_k\, k}{k_0}+\f{3}{\alpha}
-\f{3\,i\,k_0}{\alpha^2\, z_k\, k}\r)
{\rm e}^{-i\,\alpha\, z_k\, k/k_0}.
\end{eqnarray}
\end{subequations}
Note that we have arrived at these expressions for $A_k^{(\gamma)}$ and
$B_k^{(\gamma)}$ by matching the solution $f_k^{({\rm II})}$ [cf. 
Eq.~(\ref{eq:mfk-1})] and its derivative with the corresponding quantities 
in the first domain (\ie~$f_k^{({\rm I})}$ and its derivative) at 
$\eta=-\alpha\,\eta_0$.
We evaluate the tensor power spectrum after the bounce at $\eta=\beta\,\eta_0$ 
(with $\beta=10^2$), as we had done earlier.
It can be expressed as
\begin{equation}
\label{tps-fk}
{\cal P}_{_{\rm T}}^{(\gamma)}(k)
=\f{8}{\Mpl^2}\f{k^3}{2\,\pi^2}\,
\vert A_k^{(\gamma)}+B_k^{(\gamma)}\,g(\beta)\vert^2.
\end{equation}

\par

\begin{figure}[!t]
\begin{center}
\includegraphics[width=10.00cm]{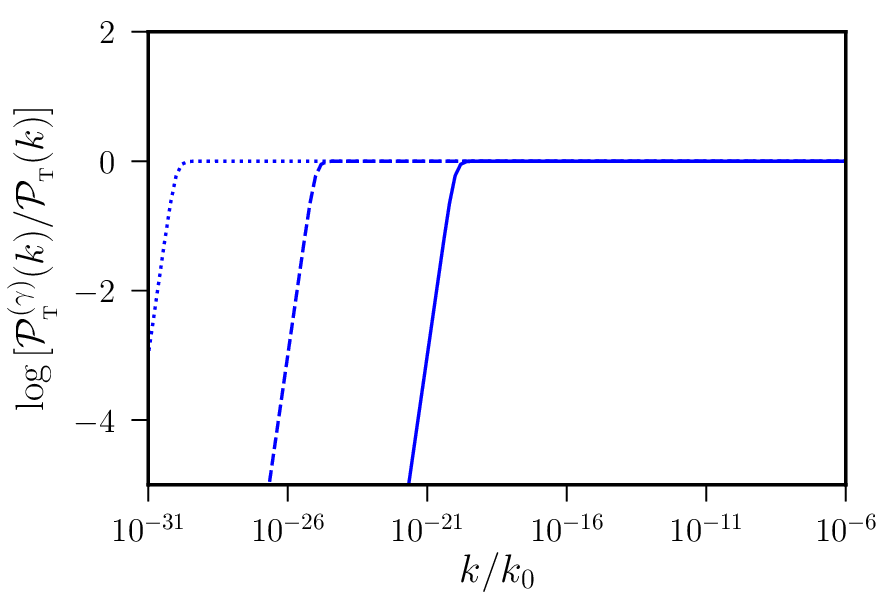}
\caption{The logarithm of the ratio of the CSL modified tensor power 
spectrum to the standard power spectrum has been plotted as a 
function of $k/k_0$ for the matter bounce ($q=1$) scenario.
We have set $k_0/(a_0\,\Mpl)=10^{-5}$, $\alpha=10^5$ and $\beta=10^2$
in plotting these figures.
The solid, dashed and dotted lines correspond to $\sqrt{\gamma}/k_0$ 
of $10^{-20},~10^{-25}$ and $10^{-30}$, respectively. 
Note that the introduction of a CSL parameter $\gamma$ leads to a 
suppression of power in the power spectrum at large scales. 
In the suppressed part, the power spectrum behaves as~$k^3$, which 
is similar to what occurs in the case of 
inflation~\cite{jerome-2012}.}\label{fig:tps-csl}
\end{center}
\end{figure}

In Fig.~\ref{fig:tps-csl} we have plotted logarithm of the ratio of the CSL 
modified power spectrum to the unmodified power spectrum, \ie 
$\log\, [{\cal P}_{_{\rm T}}^{(\gamma)}(k)/{\cal P}_{_{\rm T}}(k)]$, as a 
function of $k/k_0$, for the same set of parameters we have worked with
earlier and for a few different choices of $\sqrt{\gamma}/k_0$. 
It is evident from the figure that, just as in the case of 
inflation~\cite{jerome-2012}, the effect of CSL on the power 
spectrum in the matter bounce is to suppress its power at 
large scales. 
We find that the power spectrum behaves as $k^3$ in its suppressed 
part, exactly as observed in inflation~\cite{jerome-2012}. 
We also note that, larger the value of the dimensionless parameter 
$\sqrt{\gamma}/k_0$, smaller is the scale at which the power gets 
suppressed. 
Since, the scales of cosmological interest lie in the range $k/k_0 
\simeq 10^{-30}$--$10^{-25}$, if we demand a nearly scale invariant 
power spectrum for the tensor modes, the value of the dimensionless 
parameter $\sqrt{\gamma}/k_0$ is constrained to be $\sqrt{\gamma}
/k_0\lesssim 10^{-30}$.


\section{Tensor power spectrum in a generic bouncing model}\label{sec:i-csl-tps-gb}

Until now, our discussions have focused on the particular case of the 
classical bounce referred to as the matter bounce scenario described 
by the scale factor~(\ref{eq:sf}), with~$q$ set to unity. 
In this section, we shall turn our attention to a more generic case where~$q$ 
is any positive real number. 
As we had mentioned at the beginning of Sec.~\ref{sec:tps-mb}, 
such symmetric near-matter bounces can be achieved with the help of two fluids
or two scalar fields. 
A specific value of $q$ can be achieved by suitably choosing the effective 
equation of states of the two sources~\cite{mb,rathul-2018,rathul-2019}.
For instance, one finds that two fluids or scalar fields with equations of 
state $w_1 =(1-q)/(3\,q)$ and $w_2 =(2-q)/(3\,q)$ lead to the scale
factor~(\ref{eq:sf}).
In order to arrive at the tensor power spectrum in these models, our approach 
would be the same as in Sec.~\ref{sec:tps-mb}, \viz~to solve Eq.~(\ref{eq:uk-eom}) 
to obtain $f_k$ and then evaluate the power spectrum using the 
definition~(\ref{eq:tps-d}).


\subsection{Tensor modes and power spectrum}\label{subsec:tps-gb}

Following the discussion in Sec.~\ref{sec:tps-mb}, we obtain the tensor modes 
$f_k$ by dividing the time of interest into two domains and working
under suitable approximations.
In the first domain, \ie~over $-\infty<\eta<-\alpha\, \eta_0$, where $\alpha\gg 
1$, the scale factor simplifies to the power law form $a(\eta)\simeq
a_0\,(k_0\,\eta)^{2\,q}$. 
In such a case, the differential equation~(\ref{eq:uk-eom}) reduces to 
\begin{equation}
f_k''+\l[k^2-\f{2\,q\,(2\,q-1)}{\eta^2}\r]\,f_k=0,
\end{equation}
and it is well known that the corresponding solutions can be written in 
terms of Bessel functions.
Upon imposing the Bunch-Davies initial conditions at very early times, we 
obtain the modes to be
\begin{eqnarray}
f_k^{({\rm I})}(\eta)
&=&\f{i}{2}\,\sqrt{\f{\pi}{k}}\,
\f{{\rm e}^{-i\,p\,\pi}}{{\rm sin}\,(n\,\pi)}\,(-k\,\eta)^{1/2}
\l[J_{-n}(-k\,\eta)-{\rm e}^{i\,n\,\pi}\,J_{n}(-k\,\eta)\r],
\end{eqnarray}
where $n=2\,q-1/2$, while $J_n(z)$ is the Bessel function of first 
kind~\cite{gradshteyn-2007}.

\par

\begin{figure}[!t]
\centering
\includegraphics[width=8.50cm]{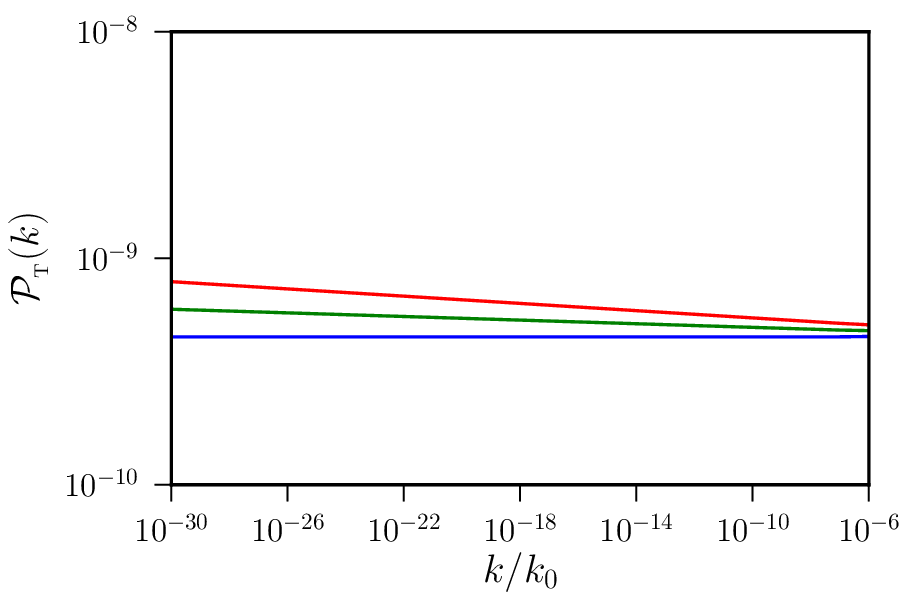}
\caption{The tensor power spectra in bouncing models corresponding to 
$q=1$ (in blue), $q=1.001$ (in green) and $q=1.002$ (in red) have been 
plotted as a function of $k/k_0$. 
We have set $k_0/(a_0\,\Mpl)=10^{-5}$, $\alpha=10^5$ and $\beta=10^2$
as in the previous figure. 
Note that the spectrum exhibits a red tilt for $q>1$.}
\label{fig:tps-p}
\end{figure}

In the second domain, \ie~over $-\alpha\,\eta_0<\eta<\beta\,\eta_0$, we 
can ignore the $k^2$ term in Eq.~(\ref{eq:uk-eom}) for reasons discussed
earlier.
For any arbitrary value of the parameter $q$, upon integrating the resulting
equation, we find that we can express the modes $f_k$ in the domain as
follows:
\begin{equation}
f_k^{(\rm II)}(\eta)
=a(\eta)\,\l[C_k+D_k\, {\tilde g}(k_0\,\eta)\r],
\end{equation}
where the function ${\tilde g}(x)$ is given in terms of the 
hypergeometric function ${}_2F_1[a,b;c;z]$ as 
\begin{equation}
{\tilde g}(x)=x\,{}_2F_1\l[2\,q,\tfrac{1}{2};\tfrac{3}{2};-x^2\r].\label{eq:gt}
\end{equation}
The constants $C_k$ and $D_k$  are obtained by matching the solutions
in the two domains and their derivatives at $\eta=-\alpha\,\eta_0$.
They can be determined to be
\begin{subequations}
\begin{eqnarray}
C_k &=&\f{i}{2\, a_0\,\alpha^n}\,
\sqrt{\f{\pi}{k_0}}\,
\f{{\rm e}^{-i\,q\,\pi}}{\sin\,(n\,\pi)}\nn\\
& &\times\,\l[J_{-n}(\alpha\, k/k_0)-{\rm e}^{i\,n\,\pi}\,J_{n}(\alpha\, k/k_0)\r]
+D_k\, {\tilde g}(\alpha),\\
D_k &=&-\f{i}{2\,a_0\,\alpha^n}
\l(\f{k}{k_0}\r)\,\sqrt{\f{\pi}{k_0}}\,
\f{{\rm e}^{-i\,q\,\pi}}{\sin\,(n\,\pi)}
\l(1+\alpha^2\r)^{2\,q}\nn\\
& &\times\,\l[J_{-(n+1)}(\alpha\, k/k_0)
+{\rm e}^{i\,n\,\pi}\,J_{n+1}(\alpha\, k/k_0)\r].
\end{eqnarray}
\end{subequations}
Upon using these expressions in the definition~(\ref{eq:tps-d}) of the tensor
power spectrum and evaluating the spectrum after the bounce at $\eta=\beta/k_0$, 
we obtain that 
\begin{equation}
{\cal P}_{_{\rm T}}(k)=\f{8}{M_{\rm Pl}^2}\,\f{k^3}{2\pi^2}\,
\vert C_k+D_k\,{\tilde g}(\beta)\vert^2.
\end{equation}
In Fig.~\ref{fig:tps-p} we have plotted the tensor power spectrum for a set of 
values $q$ as a function of $k/k_0$ for a specific choice of parameters. 
As expected, deviations from $q=1$ introduces a tilt in the tensor power 
spectrum.
It is useful to note that the power spectrum is red tilted for $q>1$, as 
one would expect in inflation.


\subsection{Imprints of CSL}\label{subsec:-i-csl-tps-gb}

We can now readily compute the effect of CSL mechanism on the tensor 
power spectrum in a more generic classical bouncing scenario.
In order to calculate the tensor power spectrum, we need to solve the 
differential equation~(\ref{eq:m-uk-eom}) governing $f_k$ in the 
presence of CSL mechanism, which effectively replaces $k^2$ in the
governing equation by $k^2 - i\, \gamma$. 
All our previous arguments go through for a general $q$ and hence
we shall quickly present the essential results.

\par

\begin{figure}[!t]
\begin{center}
\includegraphics[width=8.50cm]{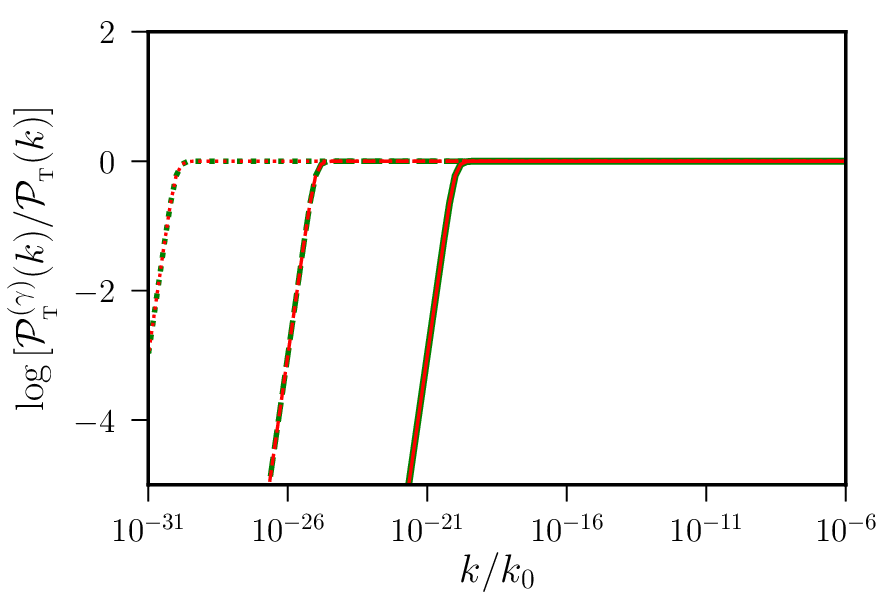}
\caption{The behavior of 
$\log\, [{\cal P}_{_{\rm T}}^{(\gamma)}(k)/{\cal P}_{_{\rm T}}(k)]$ 
has been plotted as a function of $k/k_0$ for a bounce with scale 
factor described by the indices $q=1.001$ (in green) and for $q=1.002$ 
(in red). 
We have worked with the same values of $k_0/(a_0\,\Mpl)$, $\alpha$ and 
$\beta$ as in the earlier figures. 
The solid, dashed and dotted curves correspond to $\sqrt{\gamma}/k_0$ 
of $10^{-20}, 10^{-25}$ and $~10^{-30}$, respectively. 
This figure clearly shows that the CSL mechanism leads to a suppression 
of power at large scales regardless of the value of~$q$.}\label{fig:tps-p-gamma}
\end{center}
\end{figure}

We find that the CSL modified tensor mode in the first domain is given by
\begin{eqnarray}
f_k^{({\rm I})}(\eta)
&=&\f{i}{2}\,\sqrt{\f{\pi}{z_k\,k}}\,
\f{{\rm e}^{-i\,q\,\pi}}{{\rm sin}(n\,\pi)}\;
\f{1}{\sqrt{-z_k\, k\,\eta}}
\times\,\l[J_{-n}(-z_k\, k\,\eta)-{\rm e}^{i\,n\,\pi}\,
J_{n}(-z_k\, k\,\eta)\r],\quad
\end{eqnarray}
where $n$ and $z_k$ are given by the same expressions as before.
Similarly, in the second domain, the CSL modified tensor mode can be found 
to be
\begin{equation}
f_k^{(\rm II)}(\eta)
=a(\eta)\,\l[C_k^{(\gamma)}+D_k^{(\gamma)}\, {\tilde g}(k_0\eta)\r],
\end{equation}
where, as earlier, ${\tilde g}(x)$ is given by Eq.~(\ref{eq:gt}),
while the quantities $C_k^{(\gamma)}$ and $D_k^{(\gamma)}$ are
given by
\begin{subequations}
\begin{eqnarray}
C_k^{(\gamma)}
&=&\f{i}{2\,a_0\, \alpha^n}\,\sqrt{\f{\pi}{k_0}}\,
\f{{\rm e}^{-i\,q\,\pi}}{\sin\,(n\,\pi)}
\l[J_{-n}(\alpha\, z_k\, k/k_0)
-{\rm e}^{i\,n\,\pi}J_{n}(\alpha\, z_k\, k/k_0)\r]\nn\\
&&+\,D_k^{(\gamma)}\, {\tilde g}(\alpha),\\
D_k^{(\gamma)}
&=&-\f{i}{2\,a_0\,\alpha^n}\,
\l(\f{z_k\, k}{k_0}\r)\,\sqrt{\f{\pi}{k_0}}\,
\f{{\rm e}^{-i\,q\,\pi}}{\sin\,(n\,\pi)}\,
\l(1+\alpha^2\r)^{2\,q}\nn\\
& &\times\,\l[J_{-(n+1)}(\alpha\, z_k\, k/k_0)
+{\rm e}^{i\,n\,\pi}\,J_{n+1}(\alpha\, z_k\, k/k_0)\r].
\end{eqnarray}
\end{subequations}
The resulting spectrum evaluated after the bounce at $\eta=\beta/k_0$ is 
given by
\begin{equation}
{\cal P}_{_{\rm T}}^{(\gamma)}(k)
=\f{8}{M_{\rm Pl}^2}\,\f{k^3}{2\,\pi^2}\,
\vert C_k^{(\gamma)}+D_k^{(\gamma)}\,{\tilde g}(\beta)\vert^2.
\end{equation}
As in the case of the matter bounce, in Fig.~\ref{fig:tps-p-gamma}, we have 
plotted the logarithm of the ratio of the CSL modified tensor power spectrum 
to the unmodified power spectrum, \ie the quantity 
$\log\, [{\cal P}_{_{\rm T}}^{(\gamma)}(k)
/{\cal P}_{_{\rm T}}(k)]$, for two different values of $q$ and different 
values of $\sqrt{\gamma}/k_0$. 
It is clear from the figure that the behavior of the power spectrum and the
corresponding conclusions are the same as we had arrived at in the case of
the matter bounce.

 
\section{Discussion}\label{sec:csl-c}

Generation of perturbations from quantum fluctuations in the early universe 
and their evolution leading to anisotropies in the CMB and inhomogeneities 
in the LSS provides a wonderful avenue to understand physics at the interface 
of quantum mechanics and gravitation. 
One such fundamental issue that has to be addressed is the mechanism by which 
the quantum perturbations reduce to being described in terms of classical 
stochastic variables.
In this work, we have investigated the quantum-to-classical transition of 
primordial quantum perturbations in the context of classical bouncing 
universes. 
Focusing on classical matter and near-matter bounces~\cite{mb,rathul-2018,rathul-2019}, 
and following the footsteps of earlier efforts in this direction~\cite{jerome-2012}, 
we have approached this issue from two perspectives.

\par

In the first approach, we have investigated the extent of squeezing of the
quantum state associated with a tensor mode as it evolves through a 
classical bounce. 
As in the context of inflation, the extent of squeezing grows as the modes
leave the Hubble radius. 
However, in contrast to inflation where it can grow indefinitely (depending
on the duration of inflation), we had found that the squeezing parameter 
reaches a maximum at the bounce and begins to decrease thereafter.
We had found that this behavior is also reflected in the Wigner function.

\par

Secondly, we had treated this issue as a quantum measurement problem set 
in the cosmological context, \ie we had investigated the effects of the
collapse of the original quantum state of the perturbations.
An approach which has been proposed to achieve such a collapse is the CSL 
model. 
Using the model, we had examined if the tensor power spectra are modified
due to the collapse in a class of bouncing universes. 
We had found that the CSL mechanism leads to a suppression of the tensor power 
spectra at large scales, in a manner exactly similar to what occurs in the
inflationary context. 
This suppression leads to a constraint on the collapse 
parameter.
 
\par
 
The investigation that we have carried out here shows that classical
bouncing scenarios also lead to constraints on the collapse parameter~$\gamma$ 
in a manner similar to those arrived in the context of inflation~\cite{jerome-2012}. 
However, a few clarifying remarks are in order. 
First, a fully relativistic version of CSL remains to be developed. 
This is a work in progress~\cite{Tumulka:2005ki,Bedingham:2010hz,Bedingham:2011}.
Hence, in this work, we have applied CSL, with some assumptions, to the tensor
perturbations evolving in classical bouncing scenarios. 
Currently, there is no unique way to implement CSL and different 
implementations might lead to different constraints on the collapse 
parameter\cite{Bengochea:2020efe}.  
Second, a key aspect in the implementation of CSL is the choice of the 
collapse operator.
In the context of inflation, two choices have been explored. 
One is to choose, as we have done, that the collapse operator depends on the 
Mukhanov-Sasaki variable and the other option is to choose that it depends on 
the coarse grained density contrast. 
Using the second assumption, constraints have been arrived on the collapse
parameters and it has been concluded that a particular implementation of 
CSL may be ruled out~\cite{Martin:2019jye,Martin:2019oqq}.
However, it should be underlined that these constraints do not rule out the
version of CSL we have considered here~\cite{jerome-2012,suratna-2013}. 
Moreover, without additional assumptions, it is not possible to favour one 
version of CSL over the other (in this context, see the discussion in 
Ref.~\cite{Martin:2019oqq}). 
Further, there is a priori no reason why constraints on collapse 
arrived at in the context of inflation will apply in a classical 
bouncing scenario. 
Thus, it seems important that we explore various implementations of CSL 
in the classical bouncing scenarios as well. 
Finally, in this work, we have focused on classical matter and 
near-matter bounces. 
It would be interesting to explore the mechanism in a wider class of 
bouncing scenarios.
We are currently investigating some of these issues.


\section*{Acknowledgements}

We would like to thank Debika Chowdhury for her comments on the manuscript
and Rathul Nath Raveendran for discussions. L. S. wishes to thank the Indian Institute of Technology Madras, Chennai, India, for support through Exploratory Research Project No. PHY/17-18/874/RFER/LSRI.
VS would also like to acknowledge support from NSF Grant No.~PHY-1403943.


\end{document}